\documentclass{iopart}

\usepackage[]{natbib}

\newcommand\apj{ApJ}% 
\newcommand\pasp{PASP}% 
\newcommand\aap{A\&A}%
\newcommand\mnras{MNRAS}%

\usepackage{iopams}  
\usepackage{graphicx}

\newcommand{\degrees}{^\circ}

\begin{document}

\title[The European Pulsar Timing Array]{The European Pulsar Timing Array: current efforts and a LEAP toward the future}

\author{
R.~D.~Ferdman$^{1,2,3}$, 
R.~van Haasteren$^{4}$
C.~G.~Bassa$^3$, 
M.~Burgay$^5$, 
I.~Cognard$^{1,2}$, 
A.~Corongiu$^5$, 
N.~D'Amico$^{5,6}$, 
G.~Desvignes$^{1,2}$, 
J.~W.~T.~Hessels$^{7,8}$, 
G.~H.~Janssen$^3$, 
A.~Jessner$^9$, 
C.~Jordan$^3$, 
R.~Karuppusamy$^9$, 
E.~F.~Keane$^3$, 
M.~Kramer$^{3,9}$, 
K.~Lazaridis$^9$, 
Y.~Levin$^{10}$, 
A.~G.~Lyne$^3$, 
M.~Pilia$^{5,11}$, 
A.~Possenti$^5$, 
M.~Purver$^3$, 
B.~Stappers$^3$, 
S.~Sanidas$^3$, 
R.~Smits$^7$ and
G.~Theureau$^{1,2,12}$
}

\address{$^1$Station de Radioastronomie de Nan\c{c}ay, Observatoire de Paris, 18330 Nan\c{c}ay, France}
\address{$^2$Laboratoire de Physique et Chimie de l'Environnement et de l'Espace, Centre National de la Recherche Scientifique, F-45071 Orl\'{e}ans, Cedex 2, France}
\address{$^3$The University of Manchester, School of Physics and Astronomy, Jodrell Bank Centre for Astrophysics, Alan Turing Building, Manchester, M13 9PL, UK}
\address{$^4$Leiden Observatory, Leiden University, P.O.~Box 9513, NL-2300 RA Leiden, The Netherlands}
\address{$^5$INAF-Osservatorio di Cagliari, Ioc Poggio dei Pini, strada 54, 09012, Capoterra, Italy}
\address{$^6$Universita' di Cagliari, Dipartimento di Fisica, Monserrato, Italy}
\address{$^7$Netherlands Institute for Radio Astronomy (ASTRON), Postbus 2, 7990 AA Dwingeloo, The Netherlands}
\address{$^8$Astronomical Institute ``Anton Pannekoek'', University of Amsterdam, 1098 SJ Amsterdam, Netherlands}
\address{$^9$Max-Planck-Institut f\"ur Radioastronomie, Auf dem H\"ugel 69, 53121 Bonn, Germany}
\address{$^{10}$Loretz Institute, Leiden University, P.O.~Box 9506, NL-2300 RA Leiden, The Netherlands}
\address{$^{11}$Universita' dell'Insubria, Dipartimento di Fisica, Via Valleggio 11, I-22100 Como, Italy}
\address{$^{12}$GEPI, Observatoire de Paris, Centre National de la Recherche Scientifique, Universit\'{e} Paris Diderot, 92195 Meudon, France}

\ead{robert.ferdman@manchester.ac.uk}

\begin{abstract}
The European Pulsar Timing Array (EPTA) is a multi-institutional, multi-telescope collaboration, with the goal of using high-precision pulsar timing to directly detect gravitational waves.  In this article we discuss the EPTA member telescopes, current achieved timing precision, and near-future goals.  We report a preliminary upper limit to the amplitude of a gravitational wave background.   We also discuss the Large European Array for Pulsars, in which the five major European telescopes involved in pulsar timing will be combined to provide a coherent array that will give similar sensitivity to the Arecibo radio telescope, and larger sky coverage.
\end{abstract}

\submitto{\CQG}
\maketitle

\section{Introduction}
\label{sec:intro}

Pulsars are highly magnetized neutron stars (NS), the remnants of the supernovae of massive stellar progenitors \citep{gol68,pac68}.  The majority of known pulsars rotate approximately once every second, and most are visible via their radio emission.  It is generally accepted that radio pulsars radiate from above their magnetic poles; with each rotation of the NS, we observe a cross-section through the emission region that cuts our line of sight at Earth \citep{gol68,pac68,gj69,stu71}.  This gives rise to the ``pulses'' that lend pulsars their name.  The intrinsic periodicity of most pulsars is remarkably stable, with typical period derivatives of $\sim 10^{-15}$ s\,s$^{-1}$ \citep[see e.g.,][]{lk05}.

There is a population of so-called ``recycled'' pulsars with rotational periods of $\sim 1-100$ ms.  They are the result of a mass-accretion episode from a stellar binary companion onto the NS.  The accreted matter carries angular momentum, resulting in  the ``spin-up'' of the NS \citep{sb76,acrs82,rs82,sv82}.  These millisecond pulsars (MSPs) are observed to be more stable as a population than the mainstream pulsars, with rotational period derivatives of $\sim 10^{-20}$ s\,s$^{-1}$ \citep[see e.g.,][]{lk05}.  They also show fewer instances of intrinsic, often low-frequency noise, corresponding to instabilities which affect the timing of mainstream pulsars.  This is frequently referred to as ``timing'' or ``red'' noise, and can be detrimental to long-term timing precision \citep[e.g.,][]{antt94, vbc+09, hlml09}.  For these reasons, in addition to their $2-3$ orders of magnitude shorter rotation rate, MSPs are thus the pulsars of choice when the highest timing precision is required, as is the case for the pursuit of gravitational wave (GW) detection.

Through measurement of the pulse arrival times at Earth, we can continually improve our model ephemeris for the rotation of a pulsar.  This is done by properly accounting for any deviations to the expected arrival times due to, for example, the motion of the Earth, the proper motion of the pulsar, or the binary parameters of the pulsar.  This pulsar ``timing'' aims to successfully account for every rotation of the NS, which is referred to as ``phase connection''; this is what provides the high precision for which radio pulsar astronomy is known.  The differences between the observed and expected times of arrival (TOAs) are referred to as ``timing residuals''.  An ideal timing model will generally give residuals that are normally distributed about a mean of zero \citep[see e.g.,][]{lk05}. 

Pulsar timing has been very successful for the measurement pulsar properties, and the study of the environments around NSs, for example.  Perhaps its most celebrated results, however, are those that have tested general relativity (GR) and other theories of gravity, which we briefly outline in the next section.  The potential that pulsar timing also holds for the direct detection of GWs is discussed in section \ref{sec:gw_detection}.  

\subsection{Testing gravitational theories with pulsars}

The very first test of GR using pulsar observations involved timing of the first-discovered binary pulsar, PSR~B1913+16.  Through high-precision timing of this double-neutron-star system, the measured decay of its orbit was found to agree spectacularly well with the orbital energy loss due to quadrupolar GW radiation from that system, as predicted by GR \citep{ht75a,tw89}.  More recently, observations of the double-pulsar system PSR~J0737$-$3039A/B, in which both NSs are observed as radio pulsars, have provided the most stringent test to date of GR in the strong-field regime \citep{ksm+06}.  In addition, several pulsars have been, and continue to be, used to test various aspects of GR and other theories of gravity.  These include, for example, probing Strong Equivalence Principle violation by setting limits on phenomena predicted to exist by alternative theories of gravity.  Among these are gravitational dipole radiation, secular evolution of binary parameters due to preferred frame effects, and a variation in Newton's gravitational constant \citep[e.g.,][]{nor90, dt91, wex00, arz03, vbs+08, lwj+09}.

\subsection{Direct detection of gravitational waves}
\label{sec:gw_detection}
Although some of the tests described in the preceding subsection compare predictions of GR and other theories of gravity to pulsar observations in order to \emph{infer} the existence of GW radiation as prescribed by GR, they fall short of being able to claim a \emph{direct} detection of GWs. 
Pulsars can, however, be used to directly detect a stochastic GW background (GWB), presumably produced by supermassive black hole binaries at the centers of distant galaxies \citep[][]{rr95a, jb03} or perhaps by exotic phenomena such as cosmic strings \citep[e.g.,][]{dv05}.  This is done by using an ensemble of pulsars as arms of a very large GW detector.  The perturbing effect of a GWB passing by the Earth would be correlated amongst observations of all pulsars.  This would manifest itself as a distinct signature in the timing residuals, once all model parameters particular to the pulsar system were properly accounted for.  The cross-correlation of these residuals is predicted to show a distinct dependence on the separations between the various pulsars observed, and so a distribution of pulsars evenly spread over as much of the sky as possible is desired for such a study \citep[e.g.,][]{det79, hd83}.

A distribution of sources in a ``pulsar timing array'' (PTA) thus provides a unique laboratory for performing this experiment.  The PTA experiment is sensitive to GWs with frequencies of $\sim 10^{-9}$ Hz, since the limiting frequency corresponds to the timespan of the data sets used, typically $1-10$ years in length.  Such a detection will be complementary to those that will be made by terrestrial and space-based interferometers such as LIGO \citep{aad+92} and LISA \citep{dan00}, which are most sensitive to wave frequencies of $\sim 10-10^3$ Hz and $\sim 10^{-4}-10^{-2}$ Hz, respectively.

The expected correlated signal is very subtle, and its detection requires the highest-precision timing, with rms of timing residuals well below $1\ \mu$s.  
It is also crucial that the pulsars used for this undertaking are stable over long timescales \citep[e.g.,][]{vbc+09}.  As stated in section \ref{sec:intro}, most MSPs fit this description.  Ideally, observations of $20-40$ MSPs are required to be observed over $5-10$ years, with consistent $\sim 100$ ns rms for the resulting timing residuals, in order to be able to significantly distinguish a GWB signal in the measured cross-correlation spectrum \citep[][]{jhlm05}.

The direct detection of GWs using a PTA is certainly a very challenging task.  There now exist three major research groups dedicated to this effort: The Parkes Pulsar Timing Array (PPTA) in Australia, which uses the Parkes radio telescope \citep{vbb+10}; the North American Nanohertz Observatory for Gravitational Waves (NANOGrav) collaboration in North America, which takes its data from the Arecibo and Green Bank telescopes \citep{jfl+09}; and the European Pulsar Timing Array (EPTA).  These groups have begun to develop a partnership, with the goal of sharing resources in an International Pulsar Timing Array (IPTA) effort, which will be necessary to achieve the objective of GW detection \citep{haa+10}.  In the following section, we focus on the instruments used by the EPTA and its current progress; in section \ref{sec:leap}, we describe the Large European Array for Pulsars (LEAP), the next step for the EPTA toward GW detection; finally, we conclude in section \ref{sec:summary} with a short summary.

\section{The European Pulsar Timing Array}

\begin{table}[t]
\caption{\label{tab:tel}Comparison of telescopes taking part in PTA projects.}
\lineup
\begin{tabular*}{\textwidth}{@{}l*{15}{@{\extracolsep{0pt plus 12pt}}l}}
\br
Telescope  &Diameter &Aperture   &System          & Allocated        & Declination   \\
           &(m)      &efficiency$^{\rm a}$ &temperature$^{\rm a}$ (K) & time (h/mo)   & range ($\degrees$)        \\
\mr
\multicolumn{6}{c}{\bf EPTA member telescopes}\\
\mr
Effelsberg        &100           &  0.54  &  24  &  \024  &  $\ge -30$   \\
Lovell            &\076.2        &  0.55  &  30  &  \048  &  $\ge -35$   \\
Nan\c{c}ay        &\094$^{\rm b}$ &  0.48  &  35  &  250   &  $\ge -39$   \\
Sardinia$^{\rm c}$ &\064          &  0.60  &  25  &  \030  &  $\ge -46$   \\
WSRT              &\096$^{\rm b}$ &  0.54  &  29  &  \032  &  $\ge -30$   \\
\mr
LEAP$^{\rm c}$     &194           &  0.54  &  30  &  \024  &  $\ge -39$$^{\rm d}$   \\
\mr
\multicolumn{6}{c}{\bf Other PTA telescopes}\\
\mr
Arecibo           &305         &  0.50  &  30  &  \0\08   &  $[-1, +38]$ \\
GBT               &100         &  0.70  &  20  &  \018    &  $\ge -46$   \\
Parkes            &\064        &  0.60  &  25  &  100     &  $\le +26$   \\
\br
\end{tabular*}
$^{\rm a}$Values are for 1.4 GHz.\\
$^{\rm b}$Circular-dish equivalent diameter.\\
$^{\rm c}$First light expected in 2010; shown are projected values.\\
$^{\rm d}$The limiting declination range from combining at least two EPTA member telescopes. \\
\end{table}

The EPTA is a multi-national European collaboration of pulsar astronomers with the common aim of performing and improving high-precision pulsar timing, principally for the purpose of detecting GWs.  It is also the goal of the EPTA to discover new candidate PTA pulsars, and as such, EPTA members are involved in various international pulsar search programs.   The EPTA consists of members from various institutions in France, Germany, Italy, the Netherlands, and the United Kingdom, with continued ties to research groups in other countries around the world.

\subsection{Telescopes}
The EPTA is unique among PTA collaborations, in that it employs the data from \emph{five} large-diameter radio telescopes, all of which have ongoing pulsar observing programs, including those focused specifically on collecting MSP data for the purposes of a PTA. In what follows, we describe the key features of each of these observatories.  A summary of properties of the five telescopes described above, and a comparison with the other PTA telescopes, can be found in Table~\ref{tab:tel}.  For a direct comparison of the timing capabilities of the three PTA groups, see \citet{haa+10}.

\paragraph{\bf Effelsberg.}
The Effelsberg radio telescope, administered by the Max-Planck-Institut f\"{u}r Radioastronomie, is located in Effelsberg, Germany.  At 100 meters in diameter, it is the world's second-largest fully steerable radio telescope, and can perform observations at frequencies between 0.365 and 95.5 GHz.  For pulsar observations, it hosts a digital filterbank (DFB) system that can take data over 1 GHz of observing bandwidth.  It is also capable of performing coherent dedispersion over 112 MHz of bandwidth, and provides full polarization data, with plans to expand this observing window in the near future.

The Effelsberg telescope has recently installed a new seven-beam focal plane L-band ($\sim 1.4$ GHz) receiver.  This will most notably be used in an upcoming pulsar search campaign, scheduled to begin in December 2009, in which about 500 ``normal'' pulsars, and 100 MSPs are expected to be found. 

\paragraph{\bf Lovell.}
The 76-meter Lovell radio telescope is located at the Jodrell Bank Observatory in the United Kingdom.  Its operation is overseen by the Jodrell Bank Centre for Astrophysics at the University of Manchester.  Observing primarily at 0.35 and 1.4 GHz, it is a key member of the MERLIN and European VLBI networks.

For pulsar research, it has a DFB backend similar to that at the Effelsberg telescope, and is also capable of wide-bandwidth observations.  The addition of a coherent dedispersion instrument is also being implemented, in order to enable even higher-precision data to be taken for pulsar timing observations.

\paragraph{\bf Nan\c{c}ay.}
Located in central France, the 94-meter-diameter-equivalent Nan\c{c}ay radio telescope is the fourth-largest single-dish telescope in the world, and observes at frequencies ranging between 1.1 and 3.5 GHz.  It currently houses the Berkeley-Orl\'{e}ans-Nan\c{c}ay pulsar backend, which is currently designed to coherently dedisperse pulsar data over 128 MHz of bandwidth; this will be expanded to cover 400 MHz of observing bandwidth in the near future \citep[e.g.,][]{ctdf09}.  There is also a DFB mode that can perform observations over a similar bandwidth, and is used primarily for search observations.

Although its meridian-style design restricts the telescope to track a given source for about 1 hour, pulsar observations are heavily scheduled at the Nan\c{c}ay telescope, amounting to an average of 250 hours of telescope time per month.  This, along with its state-of-the-art instrumentation, makes this telescope very well-equipped for long-term pulsar timing observations.

\paragraph{\bf Westerbork Synthesis Radio Telescope (WSRT).}
Operated by ASTRON in the Netherlands, The WSRT is a $14\times25$-meter telescope array, whose signal is combined in phase to form the equivalent of a single 96-meter radio telescope.  The WSRT can currently observe at frequencies between 0.12 and 8.7 GHz, of which pulsar timing data is regularly collected at 0.35, 1.4, and 2.3 GHz.  The WSRT is equipped with the PuMa-II pulsar backend instrument \citep{ksv08}, which performs coherent dedispersion of the incoming pulsar signal over a bandwidth of up to 160 MHz.  A major upgrade is in the works for outfitting each of the individual telescopes with a focal-plane array receiver, called APERTIF \citep[APERture Tile In Focus;][]{vov+08}.  This will effectively give the WSRT a factor of 25 increase in field of view.  It will reduce the observable frequency window to range between $1.0-1.7$ GHz, but will allow for 300 MHz of continuous observing bandwidth.

\paragraph{\bf Sardinia.}
In Italy, the National Institute of Astrophysics (INAF; Istituto Nazionale di Astrofisica), is constructing the Sardinia Radio Telescope (SRT), a fully-steerable dish that will be 64 meters in diameter \citep{taa+08}.  Most useful for pulsar research will be its planned dual-receiver system that will be capable of performing simultaneous observations at P-band ($305-425$ MHz) and L-band ($1.3-1.8$ GHz).  At the present time it is planned that the SRT will house a DFB backend for pulsar observations. First light for the SRT is planned for 2010.

\vspace{1pc}
The EPTA is distinct among PTA collaborations, in that it continues to provide valuable experience in combining data from several observatories.  This will demonstrate how to effectively and simultaneously deal with these varied data sets \citep[see, e.g.,][]{gsk+08}.  This will be of great importance for a future combined effort of an IPTA collaboration that will combine the data and resources of the NANOGrav and PPTA groups with those of the EPTA. 

\begin{table}[tp]
\caption{\label{tab:msp}Sample of MSPs currently monitored by the EPTA.}
\lineup
\begin{tabular*}{\textwidth}{@{}l*{15}{@{\extracolsep{0pt plus 12pt}}l}}
\br
Pulsar            &Total data     &rms of  \\   
name              &time span (yr) &residuals ($\mu$s) \\
\mr
PSR~B1640+22      &12             &1.6     \\
PSR~J1713+0747    &11             &0.73    \\
PSR~J1744$-$1134  &10             &0.55    \\
PSR~B1855+09      &23             &1.7     \\
PSR~J1909$-$3744  &\04            &0.11    \\
PSR~J1918$-$0642  &\07            &2.2     \\
\br
\end{tabular*}
\end{table}

\subsection{Current progress}

All the currently operational telescopes of the EPTA have observing programs that involve the regular monitoring of MSPs.  The quality of the data, and timing precision obtained, is comparable with those of the other PTA collaborations.  Table~\ref{tab:msp} shows a sample of six MSPs currently observed by the EPTA,  their respective data set time spans, and timing residual rms values.

\begin{figure}[tp]
\begin{center}
\begin{minipage}[!h]{0.51\linewidth}
\centering
\includegraphics[width=1.3\textwidth]{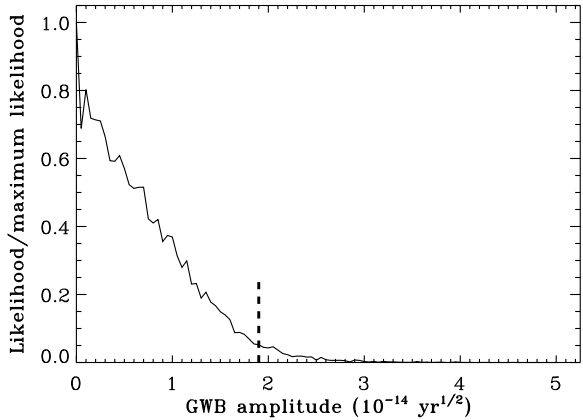}
\end{minipage}\hfill
\begin{minipage}[!h]{0.49\linewidth}
\centering
\caption{Likelihood function of the GWB amplitude derived from EPTA pulsar timing data, normalized so that the peak of the function has a value of 1.0.  The dashed vertical line shows the 95\% upper limit to the GW amplitude of $1.9\times 10^{-14}$ yr$^{-1/2}$, corresponding to a GW energy density $h_0^2 \Omega_{\rm GW} = 2.1\times 10^{-7}$ for a wave frequency of 1 yr$^{-1}$.}
\label{fig:epta_limit}
\end{minipage}
\end{center}
\end{figure}

There is also an active effort to model the observed timing residuals into a measurement of a GWB amplitude that is constrained by the EPTA data.  This involves a Bayesian statistical analysis, employing a Markov Chain Monte Carlo algorithm to sample the likelihood function of the GWB amplitude \citep{hlml09}.  At the present time, the timing precision of the EPTA pulsars is insufficient to achieve a significant GWB detection.  However, using this method, we obtain a preliminary 95\% upper limit to the GWB amplitude of $1.9\times10^{-14}$ yr$^{-1/2}$; the likelihood function derived from this analysis is shown in Figure~\ref{fig:epta_limit}.  For a wave with a frequency of of 1 yr$^{-1}$, this corresponds to a GW energy density $h_0^2 \Omega_{\rm GW} = 2.1\times 10^{-7}$ (van Haasteren et~al. 2010, in prep.).  With more data, this limit will continue to become further constrained.

\section{LEAP: The Large European Array for Pulsars}
\label{sec:leap}

The existing collaboration of EPTA member institutions, together with the close proximity of 100-meter-class radio telescopes from each other, as well as the already well-established pulsar observing programs at these telescopes, presents a unique opportunity to combine the signals from the major European observatories in a coherent array.  Such an endeavor will provide a significant boost to the quality of data used by the EPTA, enhancing its potential for detecting a GWB.  This is the primary objective of the LEAP project, which is expected to see first light by the end of 2010.  This section provides a short introduction to this project; for a more in-depth reference on LEAP, its capabilities, anticipated results, and contribution to PTA studies, see Stappers et~al (2010, in prep.).

% Need to put this figure here to ensure it appears on its own page

\begin{figure}
\begin{center}
\includegraphics[width=0.45\textwidth,angle=-90]{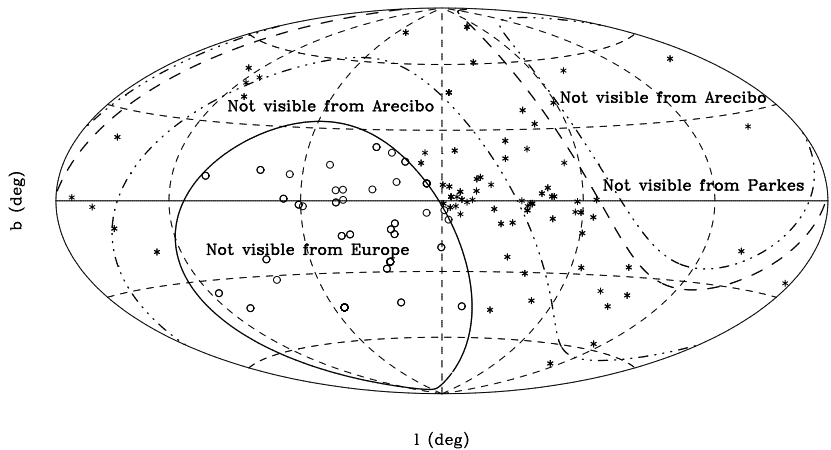}
\caption{Comparison of the sky coverage limits of the Arecibo (dash-dot-dot-dot lines) and Parkes (strong dashed lines) telescopes with the future LEAP telescope array (strong solid line).  Plotted are known MSPs, with those visible from Europe (and thus LEAP) labelled as stars.  The central vertical and horizontal solid lines represent $0\degrees$ in l and b, respectively.  Divisions of $60\degrees$ in l and $30\degrees$ in b are represented by weak dashed lines.\label{fig:leap_sky}}
\end{center}
\end{figure}

LEAP will utilize techniques adapted from existing long-baseline arrays such as VLBI \citep[e.g.,][]{dtbw07, bmg10} and MERLIN.  However, unlike the imaging data that are obtained from those collaborations,  LEAP will be designed specifically for taking pulsar timing data.  The goal is to produce a time series from all the telescopes added coherently in phase, significantly increasing the sensitivity of pulsar observations.  This will require techniques beyond the current state of the art in data-handling and processing.  The ``tied array'' formed by the EPTA member telescopes will correspond to the equivalent of a single dish of approximately 194 meters in diameter or $29\,500$ m$^2$ of collecting area, comparable to that of the illuminated Arecibo radio telescope.  This will substantially improve the timing results of the current timing array pulsars.  Perhaps more importantly, the sensitivity of LEAP will allow for the precise study of MSPs that have otherwise been too weak for PTA study with other telescopes, including the individual member observatories of the EPTA.  An important advantage that LEAP will have over the Arecibo telescope will be its much larger sky coverage.  This will allow access to a greater number of currently-known pulsars---as well as those that will be discovered in the future---that can be used as part of a PTA (see Table~\ref{tab:tel} for a comparison between telescopes).  Figure~\ref{fig:leap_sky} shows the distribution of known MSPs on the sky, and compares the ranges of sky visible from Arecibo, Parkes, and Europe (and thus LEAP).  This will make LEAP the largest fully-steerable telescope in the world, and among the most sensitive, producing ideal data for high-precision pulsar timing and a major step forward in reaching PTA goals \citep[e.g.,][]{svk09}.

The initial LEAP observing strategy will be to observe $20-30$ MSPs over a 24-hour period on an approximately monthly basis for 3 years.  This will be done over 100 MHz of bandwidth, at an observing frequency of 1.4 GHz.  The Nyquist-sampled raw voltage data will be recorded to hard disk, both from the MSPs themselves, and from calibrator sources. The latter of these data sets will provide the phase information that is needed for the multi-telescope tied-array to be achieved.  The final data-combining step will be performed at the University of Manchester, where a coherent sum of the data in the directions of the observed pulsars will be calculated using their on-site supercomputer.  The resulting high resolution, high sensitivity time-series data will then be processed by more familiar software to generate coherently-dedispersed, integrated pulse profiles.  The final step will be to calculate the pulse TOAs, which are ultimately used for the high-precision timing program to detect GWs.

However, this project will have its share of challenges; principal among these will be achieving a coherent sum of the raw recorded time-series data from all participating telescopes.  Similar procedures are used for doing this at the WSRT; however, LEAP will be composed of dishes at distances over two orders of magnitude further away from each other than those of the WSRT, each having very different properties and observing systems.  Furthermore, the tied-array will be formed in \emph{software} rather than hardware.  While this will provide greater flexibility than a system that combines the raw data in hardware, software-combining techniques are not yet sufficiently well-established for this type of data set.  There will also be a challenge in data management that must be overcome if LEAP is to be successful, as the volume of data that will be produced in a single observing session is expected to be in the hundreds of terabytes.  Other issues with which to contend will be the difficult, yet crucial tasks of ionospheric calibration, characterization of radio frequency interference, and identification of the various other systematic effects on the data that will likely arise.

All these potential difficulties will undoubtedly be encountered by future radio astronomy projects such as the Square Kilometer Array (SKA).  Thus, in addition to advancing the PTA effort to directly detect GW radiation, LEAP will provide a much-needed test-bed for the future of radio observing techniques and precise pulsar timing with combined-signal telescope arrays.

\section{Summary}
\label{sec:summary}

We have outlined in this article some of the challenges that are faced by the pulsar astronomy community in detecting a stochastic GWB, and have described the role of the EPTA within this international effort.  The EPTA, with its access to five large radio telescopes, already has the ability to provide high-quality pulsar timing data that is comparable to the other existing PTA groups.  It is also in the process of moving a step further with the construction of the LEAP telescope array, which will coherently combine the time-series data from the five major European observatories.  This will form a telescope that will be equivalent in sensitivity to the Arecibo radio telescope, but with much larger sky coverage.  This project will not only provide very high-precision pulsar data, but will also be a vital testing ground for future multi-telescope arrays, most important of which is the SKA.

\section*{Acknowledgements}
We are very grateful to all staff at the Effelsberg, Jodrell Bank, Nan\c{c}ay, and Westerbork radio telescopes for their help with observations used in this work. 
J.~W.~T.~Hessels is an Netherlands Foundation for Scientific Research (NWO) Veni Fellow. 
K.~Lazaridis was supported for this research through a stipend from the International Max Planck Research School (IMPRS) for Astronomy and Astrophysics at the Universities of Bonn and Cologne. 
Y.~Levin and R.~van Haasteren are supported by the NWO through Vidi Grant 639.042.607. 
M.~Purver is supported by a grant from the Science and Technology Facilities Council (STFC). 
S.~Sanidas is funded by a STFC Doctoral Training Account (DTA) scholarship. 
Part of this work is based on observations with the 100-m telescope of the Max-Planck-Institut f\"{u}r Radioastronomie (MPIfR) at Effelsberg. 
Access to the Lovell telescope is supported through an STFC rolling grant. 
The Nan\c{c}ay radio telescope is part of the Paris Observatory, associated with the Centre National de la Recherche Scientifique (CNRS), and partially supported by the R\'{e}gion Centre in France. 
The Sardinia Radio Telescope is a project of the Italian National Institute for Astrophysics (INAF). 
The Westerbork Synthesis Radio Telescope is operated by the Netherlands Foundation for Research in Astronomy (ASTRON) with support from the NWO.  
The LEAP project is funded through an European Reseach Council (ERC) grant to M.~Kramer.

\footnotesize
%\bibliographystyle{apjx}
%\bibliographystyle{jphysicsB}
%\bibliography{journals1,modrefs,psrrefs,crossrefs,myrefs}

%\section*{References}

\normalsize

\end{document}